\documentclass[10pt, a4paper, twocolumn]{article}
\usepackage[english]{babel} 

\usepackage{microtype} 

\usepackage{amsmath,amsfonts,amsthm} 

\usepackage[svgnames]{xcolor} 

\usepackage[hang, small, labelfont=bf, up, textfont=it]{caption} 

\usepackage{booktabs} 

\usepackage{lastpage} 

\usepackage{graphicx} 

\usepackage{enumitem} 
\setlist{noitemsep} 

\usepackage{sectsty} 
\allsectionsfont{\usefont{OT1}{phv}{b}{n}} 

\usepackage{geometry} 

\usepackage[T1]{fontenc} 
\usepackage[utf8]{inputenc} 

\usepackage{XCharter} 

\usepackage{fancyhdr} 
\fancypagestyle{firstpage}{ 
	\fancyhf{}
}

\newcommand{\authorstyle}[1]{{\large\usefont{OT1}{phv}{b}{n}\color{DarkRed}#1}} 

\newcommand{\institution}[1]{{\footnotesize\usefont{OT1}{phv}{m}{sl}\color{Black}#1}} 

\usepackage{titling} 

\newcommand{\HorRule}{\color{DarkGoldenrod}\rule{\linewidth}{1pt}} 

\pretitle{
	\vspace{-40pt} 
	\HorRule\vspace{10pt} 
	\fontsize{32}{36}\usefont{OT1}{phv}{b}{n}\selectfont 
	\color{DarkRed} 
}

\posttitle{\par\vskip 15pt} 

\preauthor{} 
\postauthor{}
\predate{}

\postdate{ 
	\vspace{5pt} 
	\par\HorRule 
	\vspace{5pt} 
}

\usepackage{lettrine} 
\usepackage{fix-cm}	

\newcommand{\initial}[1]{ 
	\lettrine[lines=3,findent=4pt,nindent=0pt]{
		\color{DarkGoldenrod}
		{#1}
	}{}%
}

\usepackage{xstring} 

\newcommand{\lettrineabstract}[1]{
	\StrLeft{#1}{1}[\firstletter] 
	\initial{\firstletter}\textbf{\StrGobbleLeft{#1}{1}} 
}

\usepackage{epsf,exscale,times}
\usepackage{graphicx}
\usepackage{amsmath}
\usepackage{fancyhdr}

\usepackage{algorithm2e}
\usepackage{amssymb}
\usepackage{csquotes}
\usepackage[backend=biber,sorting=none]{biblatex}
\usepackage{siunitx}

\renewcommand{\L}{\mathcal{L}}

\newcommand{\T}{\mathcal{T}}
\newcommand{\D}{\mathcal{D}}
\renewcommand{\phi}{\varphi}

\newcommand{\N}{\mathbb{N}}
\newcommand{\R}{\mathbb{R}}

\DeclareMathOperator{\argm}{arg\,min}
\newcommand{\argmin}[1]{\underset{#1}{\argm}\,}

\bibliography{bibliography.bib}

\title{A Load Balancing Surveillance Algorithm For Multifunctional Radar Resource Management}

\author{
	\authorstyle{Tobias M\"uller, Pascal Marquardt and Stefan Br\"uggenwirth}
	\newline\newline
	\institution{Fraunhofer Institute for High Frequency Physics and Radar Techniques\\
    Wachtberg, GERMANY\\
		email: tobias.mueller@fhr.fraunhofer.de}\\
}

\begin{document}

	\maketitle 
	
	\thispagestyle{firstpage}
	
This paper is a preprint of a paper accepted by International Radar Symposium (IRS) 2019.\\

\lettrineabstract{
For all multifunctional radar systems the allocation of resources plays an outstanding role. Many radars have low priority on surveillance tasks. In challenging situations this leads to neglecting of surveillance beams in directions where many other tasks are done. This document presents a technique that enables multifunctional radar systems to keep on scanning overloaded surveillance sectors under the condition that all sectors have a similar revisit time. Since radar resource management depends on the used system, two general configurations are considered in this paper. The focus lies on systems with a rotating antenna.
}

\section{Introduction}
\begin{figure}[ht] 
	\centering
	\def\svgwidth{0.35\textwidth}
\begingroup%
  \makeatletter%
  \providecommand\color[2][]{%
    \errmessage{(Inkscape) Color is used for the text in Inkscape, but the package 'color.sty' is not loaded}%
    \renewcommand\color[2][]{}%
  }%
  \providecommand\transparent[1]{%
    \errmessage{(Inkscape) Transparency is used (non-zero) for the text in Inkscape, but the package 'transparent.sty' is not loaded}%
    \renewcommand\transparent[1]{}%
  }%
  \providecommand\rotatebox[2]{#2}%
  \newcommand*\fsize{\dimexpr\f@size pt\relax}%
  \newcommand*\lineheight[1]{\fontsize{\fsize}{#1\fsize}\selectfont}%
  \ifx\svgwidth\undefined%
    \setlength{\unitlength}{595.27559055bp}%
    \ifx\svgscale\undefined%
      \relax%
    \else%
      \setlength{\unitlength}{\unitlength * \real{\svgscale}}%
    \fi%
  \else%
    \setlength{\unitlength}{\svgwidth}%
  \fi%
  \global\let\svgwidth\undefined%
  \global\let\svgscale\undefined%
  \makeatother%
  \begin{picture}(1,1.01428571)%
    \lineheight{1}%
    \setlength\tabcolsep{0pt}%
    \put(0,0){\includegraphics[width=\unitlength,page=1]{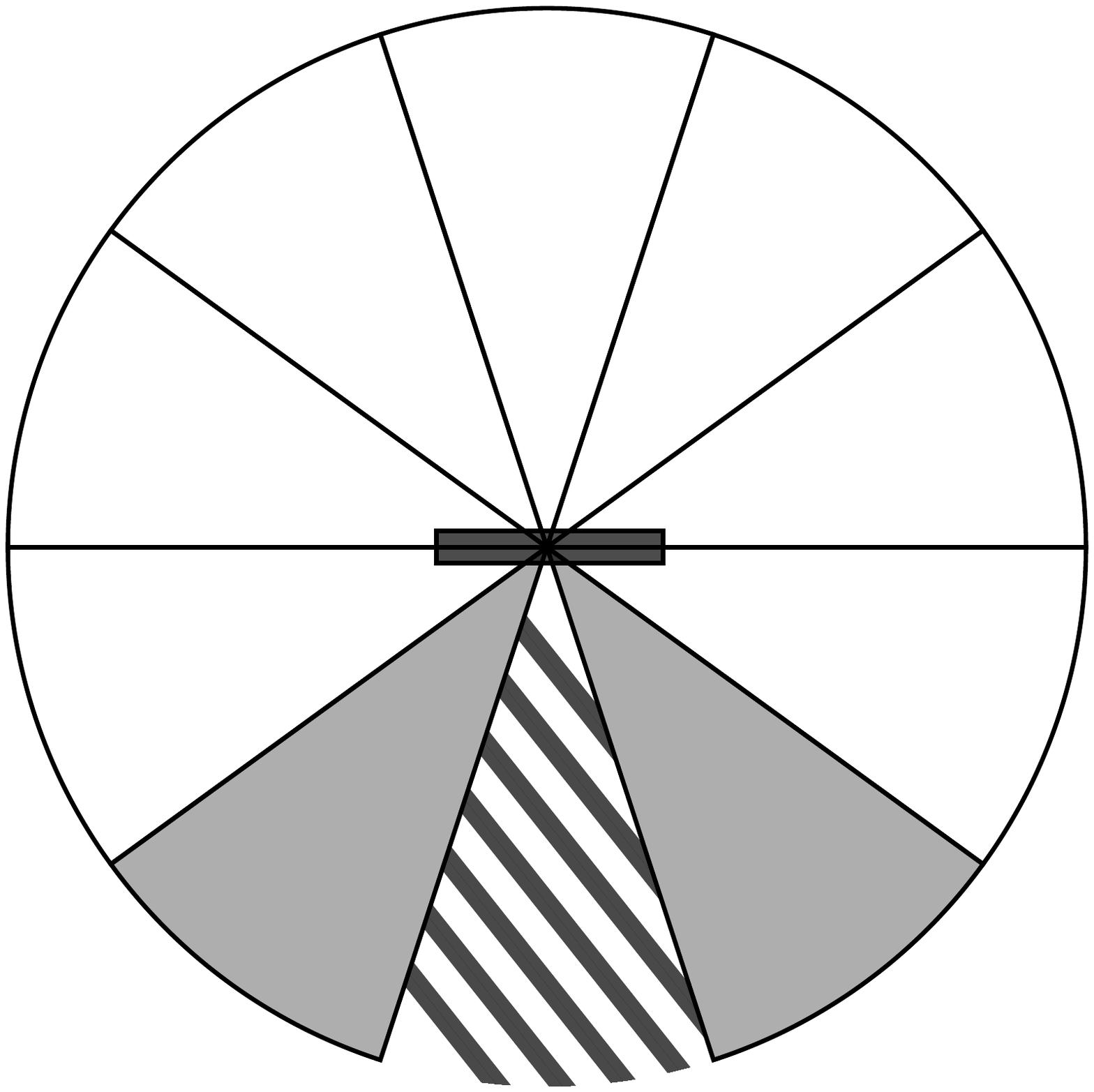}}%
    \put(0.67644646,0.60708364){\color[rgb]{0,0,0}\makebox(0,0)[lt]{\lineheight{1.25}\smash{\begin{tabular}[t]{l}Antenna\end{tabular}}}}%
    \put(0,0){\includegraphics[width=\unitlength,page=2]{FOV.pdf}}%
    \put(0.48999409,0.66435569){\color[rgb]{0,0,0}\makebox(0,0)[lt]{\lineheight{1.25}\smash{\begin{tabular}[t]{l}Rotation\end{tabular}}}}%
    \put(0,0){\includegraphics[width=\unitlength,page=3]{FOV.pdf}}%
    \put(0.54176588,0.16378973){\color[rgb]{0,0,0}\makebox(0,0)[lt]{\begin{minipage}{0.28978174\unitlength}\raggedleft FOV\end{minipage}}}%
    \put(0.44097222,0.10079369){\color[rgb]{0,0,0}\makebox(0,0)[lt]{\begin{minipage}{0.39057539\unitlength}\raggedleft Main Sector\end{minipage}}}%
    \put(0,0){\includegraphics[width=\unitlength,page=4]{FOV.pdf}}%
  \end{picture}%
\endgroup%

	\includegraphics[scale=1]{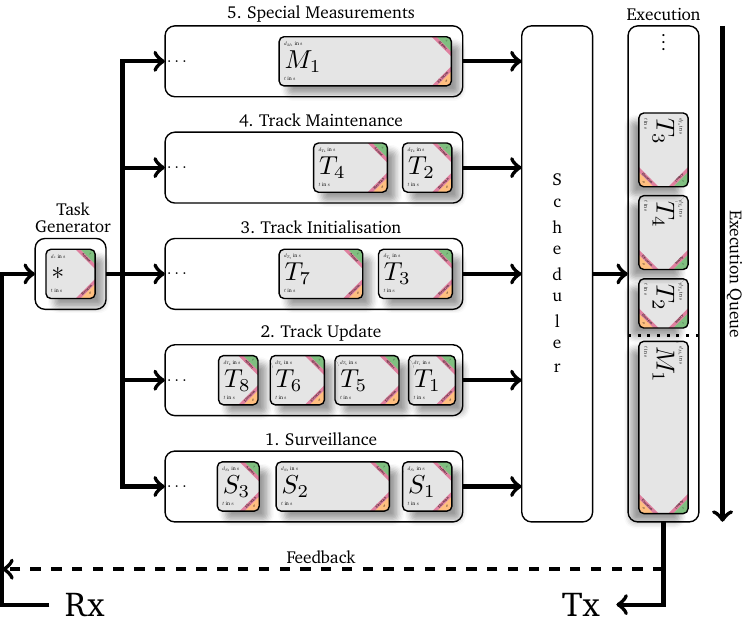}
	\caption{Generic scheduling scheme on the lower and schematic illustration of the field of view on the upper side. \label{fig:intro}}
\end{figure}

Multifunctional radar systems (MFR) offer many opportunities to adapt to the scene. This leads to the challenging task of dynamically allocating resources for different tasks such as surveillance, tracking, classification or imaging in changing environments. Due to the growing demands modern radar systems have to meet, the scheduling of surveillance tasks is treated not very much. The reason for this is that often a low priority is given to these tasks which leads to fluctuating revisit times. The left picture of Figure \ref{fig:intro} depicts a generic scheduling system. This uses queues with priorities to decide in which order tasks have to be done. The result is written into the execution queue.

This work examines the problem for surveillance resource allocation with a focus on rotating antenna systems. Since the radar system is occupied unbalanced over the area, the goal is to reach a balanced surveillance even if the available resources are unbalanced. Many publications consider a scheduling system for all radar tasks together \cite{butler97},\cite{mir07},\cite{win06} or even include prior knowledge \cite{mir06}. Investigations with focus on rotating systems are sparsely given \cite{cummings}, \cite{butlerP}. Since in many cases surveillance is given the lowest priority, highly occupied regions either lack in surveillance revisit time or are updated far away from broadside. This generally increases beam steering losses. Modern algorithms that handle this, e.g. Quality of Service \cite{hansen},\cite{charlish}, require high computational effort and an upgrade for existing systems might be complicated.
\par
The method presented in the following is computationally fast and gives the opportunity to upgrade existing systems because of its flexible setup. The result can be interpreted as a more efficient way to order the tasks within the surveillance queue than just sorting them by their deadlines (earliest deadline first \cite{edf}). It is based on dividing the surveillance space into sectors along the rotation direction. These are then interpreted as bins similar to the well known bin packing problem \cite{korte2012combinatorial}, which is NP-hard. Additionally only bins near broadside can be used to execute a surveillance task to fulfil the field of view (FOV) condition (Figure \ref{fig:intro}). This can be set smaller than the maximum possible FOV to reduce the beam steering losses.
\par
The remainder of the paper is organised as follows. In section \ref{sec:probdef} the problem is defined for the case that the available resources are known apriori. Afterwards in section \ref{sec:npcomplete} the NP-hardness of the problem is shown. Section \ref{sec:alg} describes the algorithm and in section \ref{sec:results} the results are presented and compared to an optimal solution. In section \ref{sec:othercases} the assumptions on the problem are relaxed. Section \ref{sec:conclusion} concludes the paper and gives an outlook on future work.

\section{Problem definition} \label{sec:probdef}

The surveillance volume of a radar can be described in spherical coordinates. It can be discretised in azimuth and elevation given by $(\phi,\theta) \in [0,2\pi) \times [-\pi,\pi]$. The set of all directions where a surveillance beam shall be pointed is denoted by
\begin{align}
	\L \subset \left( [0,2\pi) \times [-\pi,\pi] \right) \quad \text{and} \quad \D_\L:\L \rightarrow \R \, ,
\end{align}
where $\mathcal{D}_\L$ denotes the duration that every surveillance beam consumes. Let $N\in\N$ be a given number of surveillance sectors. This leads to the following partition of the search volume:
\begin{align}
	\L_i := &\left\{ (\phi,\theta)\in\L \quad | \quad \lfloor \frac{\phi}{2\pi}N \rfloor=i \right\} \\
	 &\text{with} \quad  \L_i \cap \L_j = \emptyset, \quad i \neq j, \quad 0 \leq i,j < N \, . \nonumber
\end{align}
In the first step the assumption is that the surveillance resources $R_i \in \R$ for every sector are known and fixed but arbitrary. 

The mechanical rotation of the antenna is considered as constant with rotation time $\varDelta t$ per sector such that a complete rotation needs $N\, \varDelta t$. W.l.o.g. the angular position starts with zero such that the position of the antenna can be written as $\phi(t)=t\frac{2\pi}{N \varDelta t}$. This position is related to the \textit{main sector} $m(t)=\lfloor \frac{\phi(t)}{2\pi}N \rfloor$. Additionally the field of view is restricted, this is here represented by a fixed number of sectors $n \in \N$ that the radar can look aside to. Therefore the \textit{set of active sectors} is defined as
\begin{align}
	F_i := \{j \in \N \mid  j = (m(i \varDelta t) + c) \, \text{ mod } N, \, |c| \leq n \}.
\end{align}

A Task $T$ is defined as a triple $(L,\D_L,i)\in\L \times \D_\L \times \N$ where the last element gives the main sector at execution time by $m(i \varDelta t)$. The components of a Task $T$ are referenced by $T_1,T_2,T_3$. The set 
\begin{align}
	\T_i := \{ T \mid T_3 \leq i \}
\end{align}
contains all tasks executed within t, i.e. $\lfloor \frac{t}{\varDelta t} \rfloor \leq i$.

\par
This leads to the following optimisation problem:
\begin{equation} \label{eq:problem}
\begin{aligned}
	 					&  {\text{minimise}} \quad t \\
						&  \text{subject to}	\\
						& \quad \forall L \in \L, \, \exists T \in \T_i \colon T_1=L,	\\
						&\quad \forall T \in \T_i:\, m(T_3 \varDelta t) \in F_{T_3},	\\
						& \sum_{
						T \in \T_i \setminus \T_{i-1}}
						T_2 < R_{m(i\varDelta t)} \quad \forall i \in \N \text{ with } i\varDelta t<t.
\end{aligned}
\end{equation}
The first condition enforces that no direction in $\L$ is neglected. The second condition enforces for all tasks to be executed within the field of view. The third condition incorporates the resource restrictions for every sector. The term $T \in \T_i \setminus \T_{i-1}$ means that only tasks that are executed within the same rotation are considered. The calculation of the current main sector $m(T_3 \varDelta t) = i$ depends on the time stamp of execution of the task $T$. 
\par
Since the focus is set to rotating antennas, this optimisation problem can be read as minimising the number of rotations that are necessary to update all surveillance beam pointing directions. It is important to see that the available resources per sector are decoupled from the mechanical rotation time since $\varDelta t$ only gives a physical upper bound for the available time. Another option is to adapt the rotation rate in a next step. Relaxing the fixed resources per sector is done in a later section. In the next section the NP-hardness of this problem will be proven. 

\section{NP-Hardness}\label{sec:npcomplete}
The NP-hardness of Problem \eqref{eq:problem} follows easily from a reduction to the bin packing problem \cite{korte2012combinatorial}. To solve bin packing with \eqref{eq:problem} the field of view can be set to $N$ such that any task can be executed on any sector. Additionally only results for $t<\varDelta t N$ are accepted. This leads to 
\begin{equation} \label{eq:binpack}
\begin{aligned}
	 					&  {\text{minimize}}	
						&&  t \\
						&  \text{subject to}
						&& \forall L \in \L, \exists T \in \T_{N \varDelta t} \colon T_1=L,	\\
						&&& \forall 0 \leq i < N: \, \sum_{T \in \T_i}
						T_2 \leq R_{i} \, .	
\end{aligned}
\end{equation}
This is already the bin packing problem.

\begin{figure}[t]
	\includegraphics[width=0.475\textwidth]{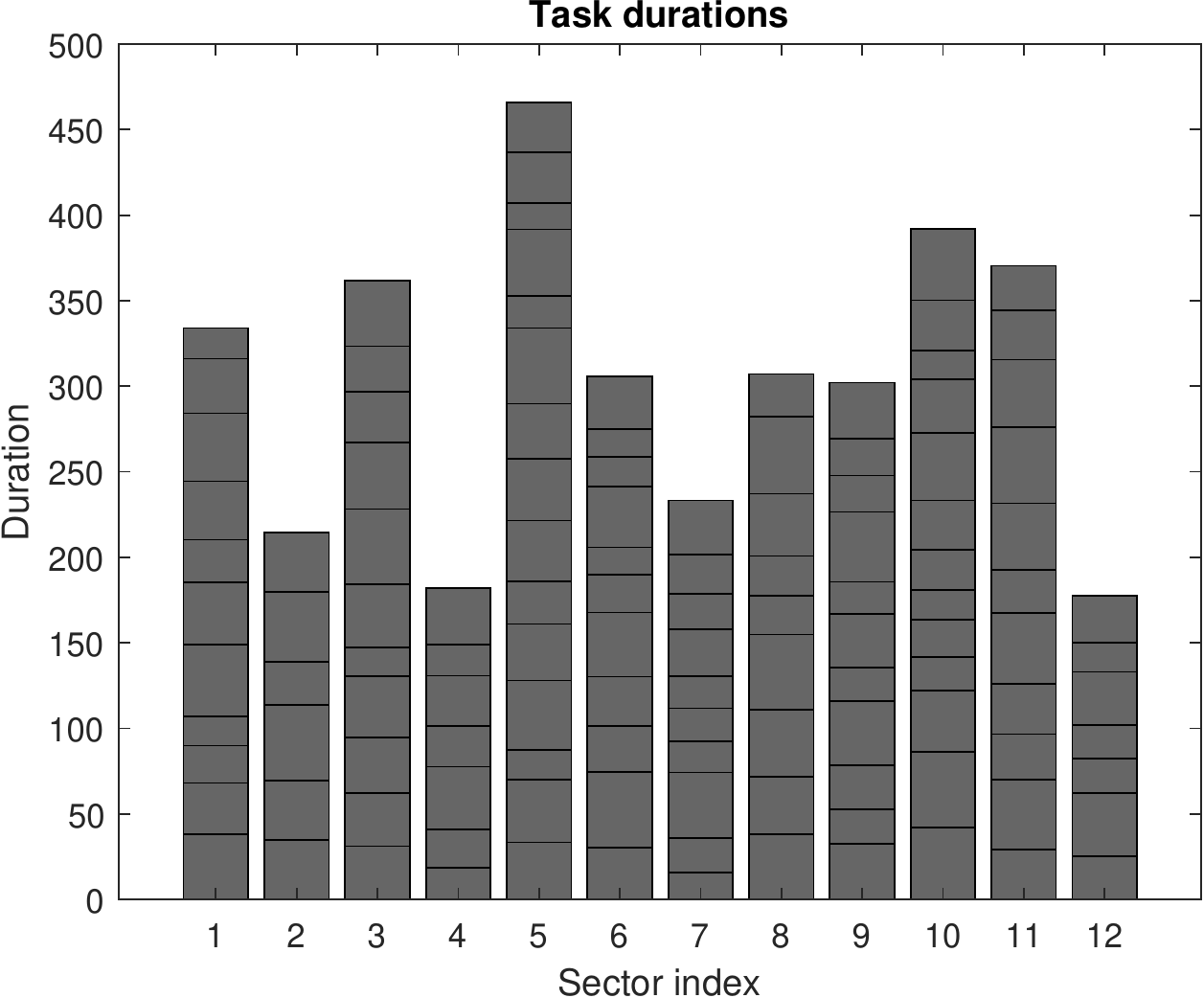} \hfill
	\includegraphics[width=0.475\textwidth]{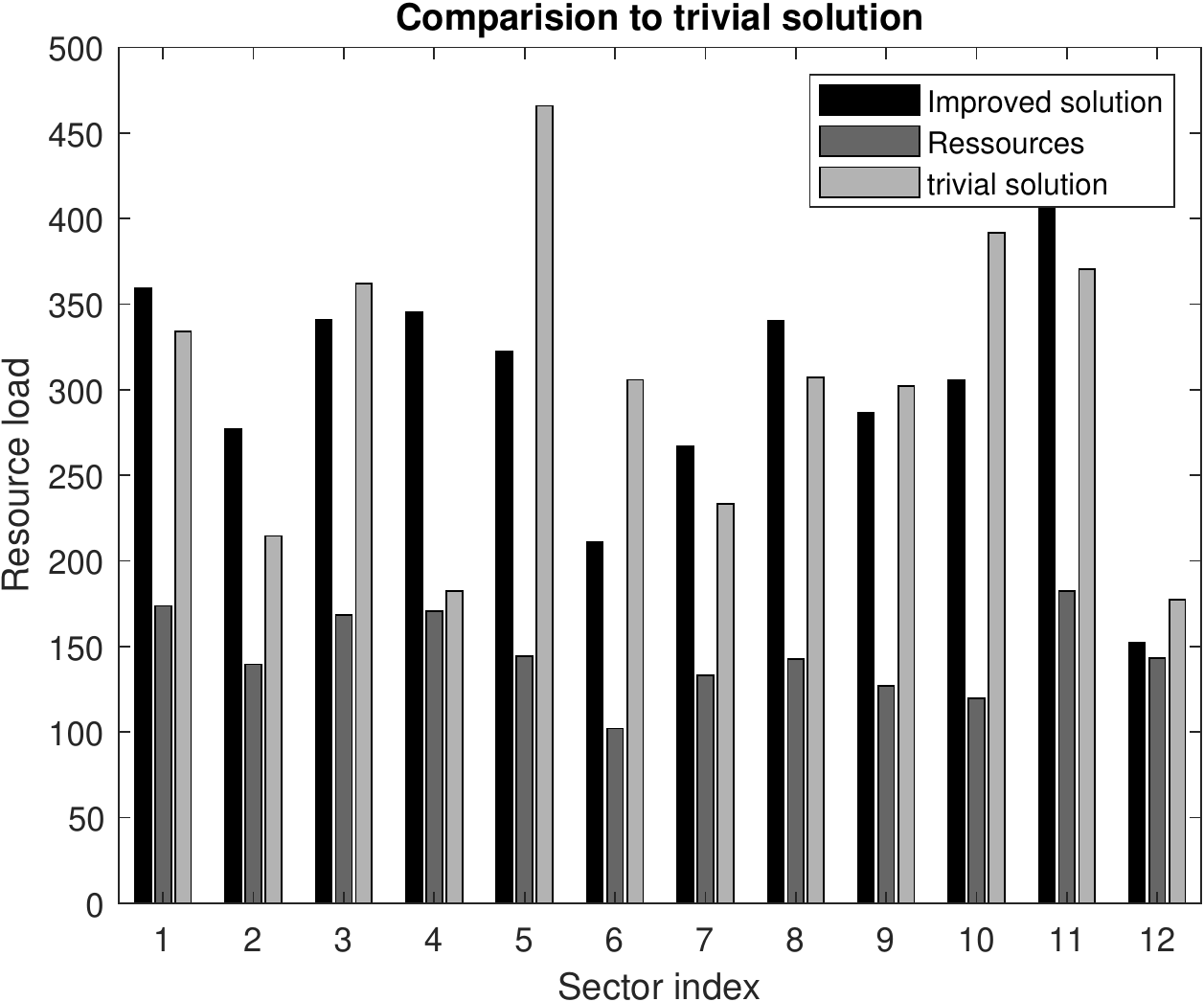}
	\caption{This example shows on the left how the different sectors are occupied, where the stacked elements indicate the durations of each surveillance update. On the right the occupancy after equalisation in comparison to the given resources is depicted.}\label{fig:example}
\end{figure}

\section{Algorithm description}\label{sec:alg}
Since the problem is NP-hard a simplified solution is presented. This is based on a greedy design and starts with a continuous calculation which makes the problem easy to solve because any portion of a task would be executable. In the next step the continuous solution is rounded off to a discrete guess. Therefore define the continuous optimal criteria:
\begin{align}
	r_{opt} &:= \frac{\sum\limits_{d \in \D_\L} d}{\sum\limits_i R_i}
\end{align}
This fraction is the absolute time demand to update all surveillance directions in relation to the total resources available per rotation. $r_{opt}$ can be used to get the optimal resource distribution per sector
\begin{align}
	r_{i_{opt}}:=r_{opt} R_i \,.
\end{align}

\begin{algorithm}

	\KwData{given $N,R,D,n$}
 	\KwResult{A partition $B_i$ on all surveillance tasks.}
 	calculate $r_{i_{opt}}$\;
 	$E=\emptyset$\;
 	\For{$i\leftarrow 1$ \KwTo $N$}{
 		$\tilde{\L}=\L_i \setminus E$ \;
 		Choose $P_1$ maximal in relation to $\tilde{\L}$ on sector i with a greedy approach\; \vspace{0.1cm}
 		
 		Calculate the sectors in field of view:\\
		$F = [i-n,\dots,i+n]$ mod $\, N$ \;
 		$\tilde{\L}=\L_F \setminus E$ \;
 		Choose $P_2 \supset P_1$ maximal in relation to $\tilde{\L}$ on sector i with a greedy approach\;

   		$B_i = P_2$\;
   		$E=E \cup B_i$\;
	}
	Distribute not used beam pointing directions:	\\
 	\While{$L \in \L \setminus E$}{
 		Set $i$ such that $L \in \L_i$\;
	 	$F = [i-n,\dots,i+n]$ mod $\, N$ \;
		$m=\argmin{j \in F} \frac{\tilde{d} \, + \sum\limits_{d \in B_j} d}{r_{j_{opt}}}$ \;
		$B_{m}=B_{m} \cup L$ \;
		$E=E\cup L$\;
	}
 \caption{Surveillance scheduling algorithm \label{alg}}
\end{algorithm}

Algorithm \ref{alg} generates a mapping from every sector to the beam pointing directions or rather their time demand. Figure \ref{fig:example} shows an example input and output of the algorithm. On the left side the update durations and their sector membership by position are shown. On the right side the available resources per rotation, and the equalised solution against the trivial solution which is just executing the task on broadside are drawn. In sector $5$ one can see the advantage in this equalisation effect.
\par
In words the algorithm chooses a maximal subset of $\L$ for every sector for which the resource allocation is lower than the optimal value $r_{i_{opt}}$. Maximal in this case means that there is no task in the given set of tasks that can be added without exceeding the threshold. So the set $P \subset \tilde{\L} \subset \L$ is \textit{maximal in relation to the set} $\tilde{\L}$ \textit{on sector} $i$ iff: 
\begin{align}
	&d^\prime+\sum_{d \in D_P} d>r_{i_{opt}}, \quad \forall d^\prime \in D_{\tilde{\L} \setminus P}	\\
	\intertext{and}
	&\sum_{d \in D_P} d \leq r_{i_{opt}} \,.
\end{align}
After that is done for all sectors there usually will be some unassociated tasks. They are associated to that sectors, where the violation of the threshold $r_{i_{opt}}$ is minimal in a relative manner.

The set $E$ is a storage for all executed tasks. The reference $r_{i_{opt}}$ is used to decide how much load a sector should handle.

The partition produced by the algorithm now defines  what surveillance tasks have to be executed depending on the current main sector. At this point the rotation rate is considered which is not used to generate the scheme itself. The decision which task will be executed next within the main sector, is then done by comparing the time of last illumination, for all $L \in B_i$ .

\section{Results} \label{sec:results}
To test the proposed algorithm, a random input is generated. The results are compared to the resource distribution without equilibration and to the optimal solution $r_{i_{opt}}$. 

\begin{figure}[t]
	\includegraphics[width=0.475\textwidth]{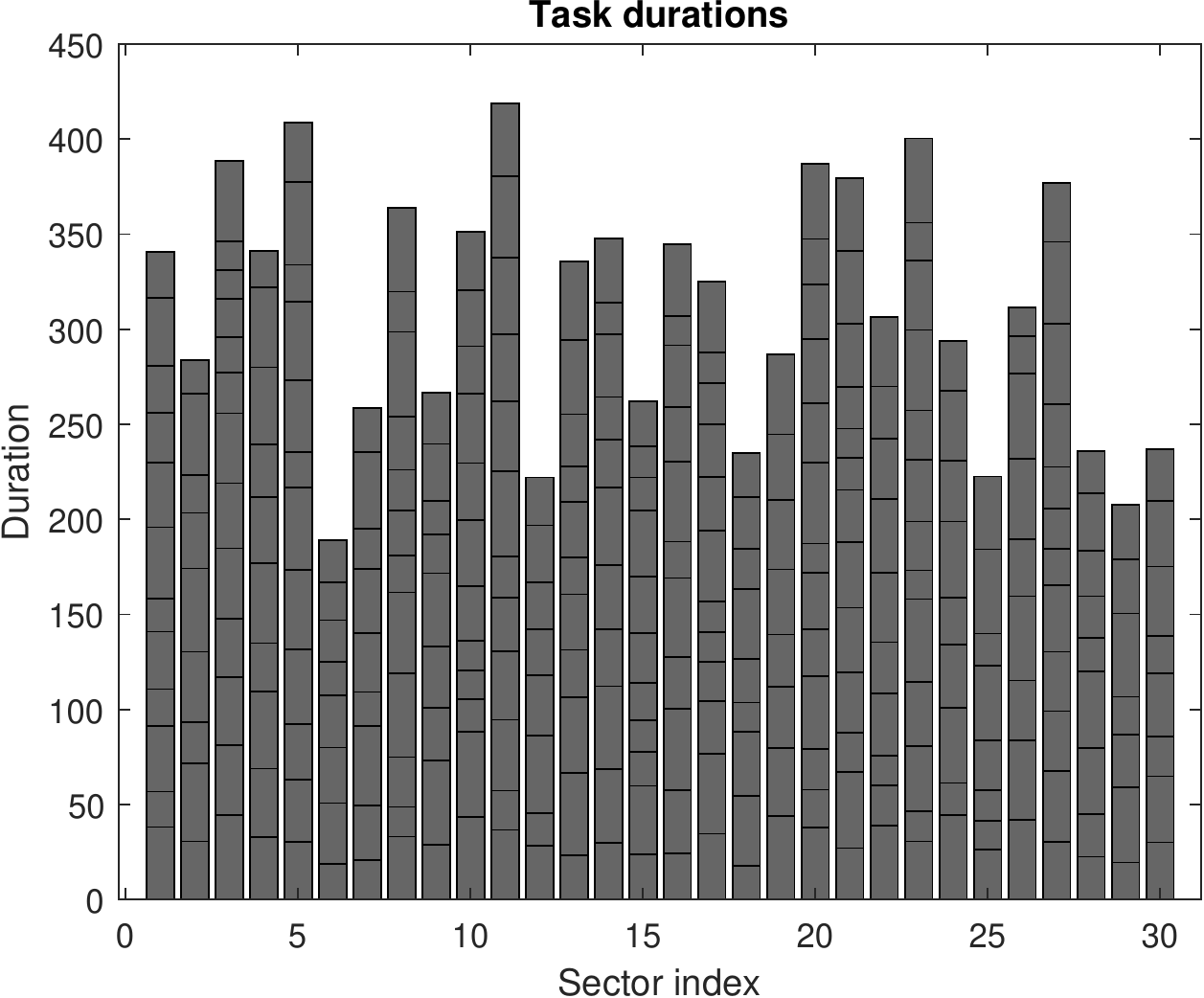}\\
	\includegraphics[width=0.475\textwidth]{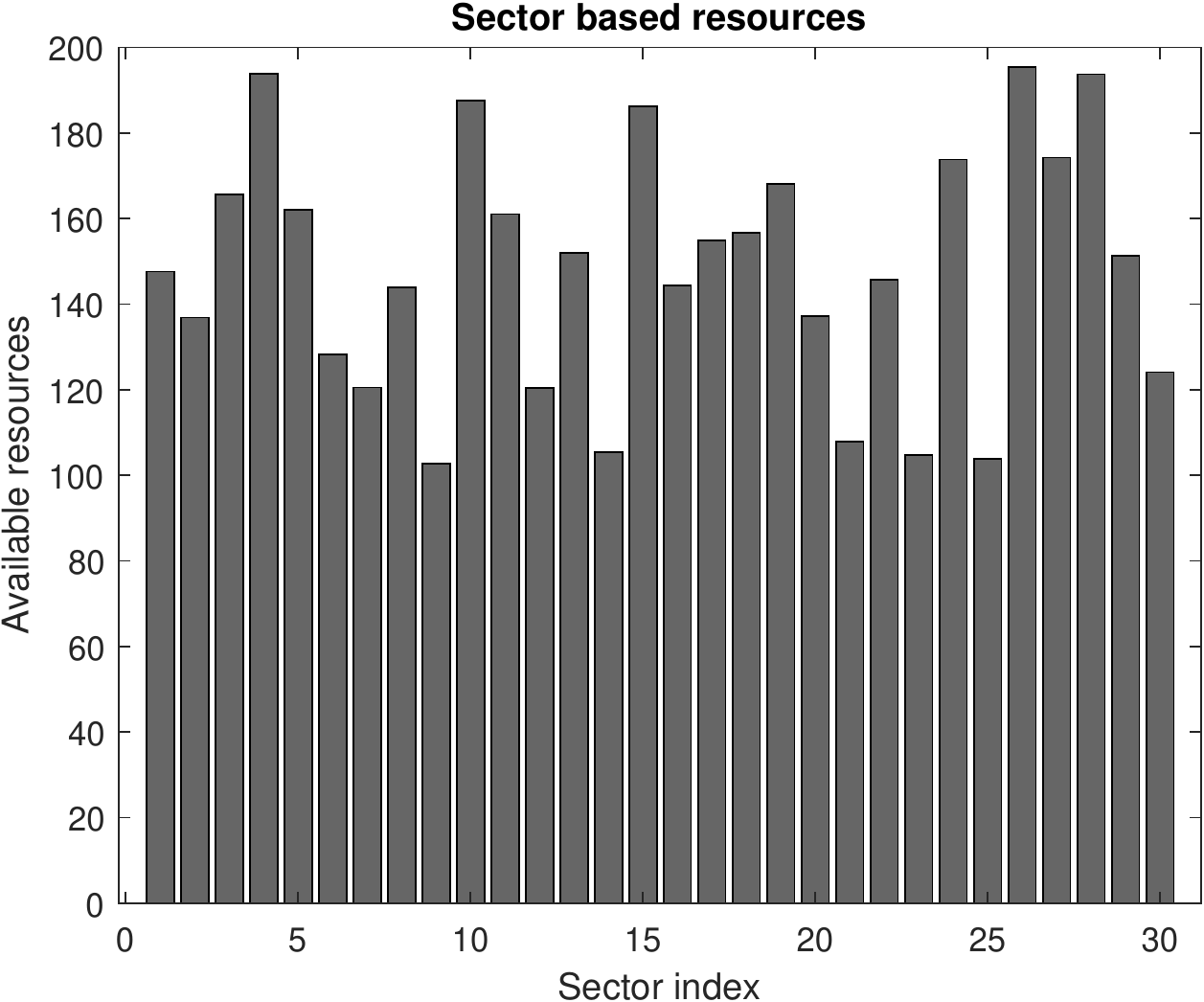}
	\caption{This experiment shows how the different sectors are occupied by the durations of their tasks (upper) and their resources (lower). The field of view is defined by $n=5$. \label{fig:ex}}
\end{figure}

\begin{figure}[t]
	\includegraphics[width=0.475\textwidth]{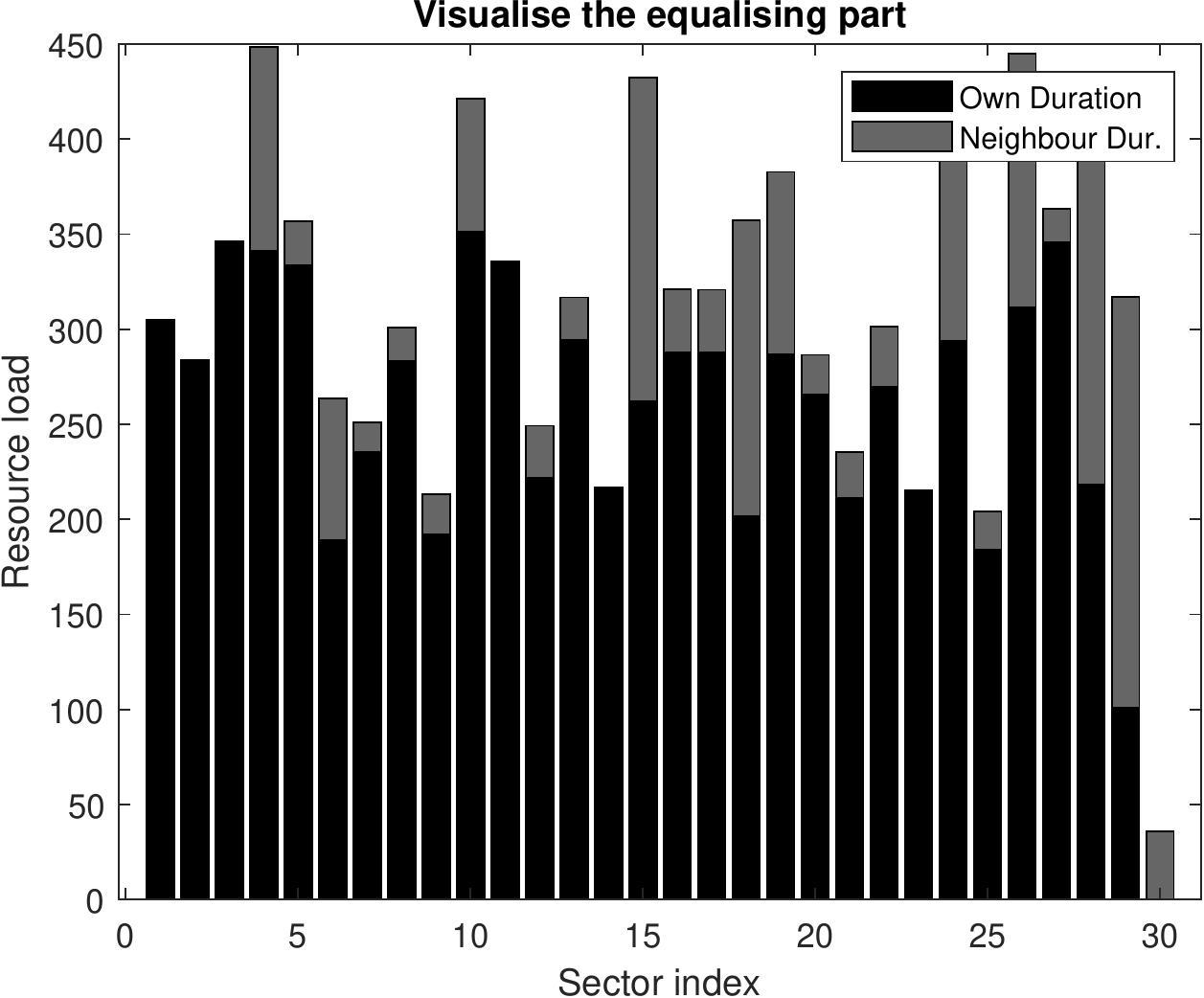}\\
	\includegraphics[width=0.475\textwidth]{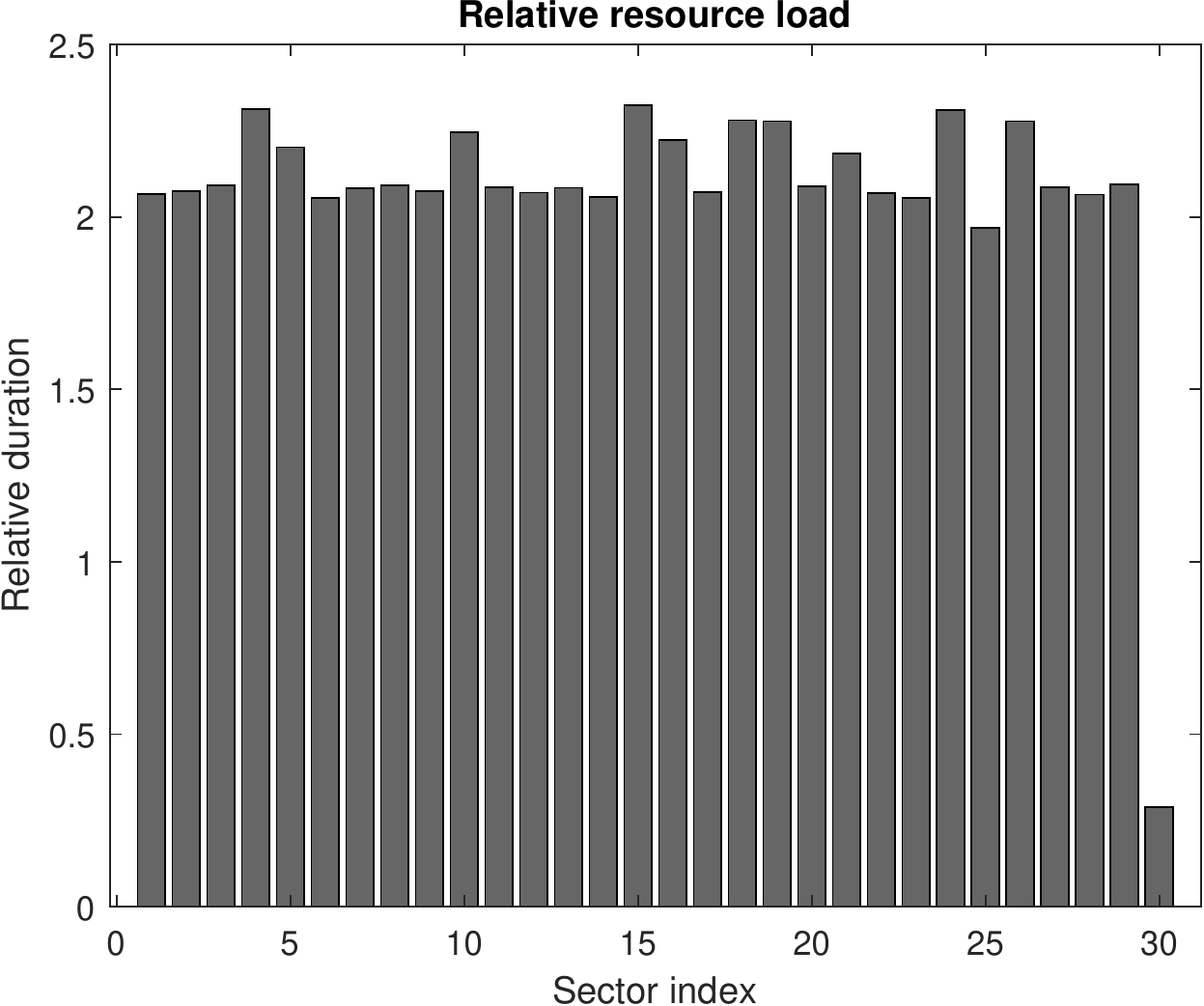}
	\caption{These graphs represent the result of the algorithm for the input shown in Figure \ref{fig:ex}. The upper shows the result after equalisation. The portion of tasks that are executed on broadside are compared to those tasks that are executed away from broadside. The lower shows the attained relative load on each sector.\label{fig:exe}}
\end{figure}

Figure \ref{fig:ex} shows an example input with $30$ sectors and a field of view of $11$ sectors ($n=5$) which is about $130^\circ$. This is a realistic standard scenario for many cases. The upper shows the update directions and their durations. The lower shows the available resources per sector. The output is shown in Figure \ref{fig:exe} where the upper shows which tasks are executed within their sector and which were used for equalisation. In sector $30$ all own tasks are  executed by other sectors such that only tasks from neighbouring sectors are left. The reason for this is that the equalisation spreads the resources on all sectors such that there is no task left in FOV for this sector to associate. An additional shifting step would be necessary to overcome this problem. Since this is an underload case, the worst case revisit time is not directly increased. The overall performance and the optimal continuous solution are similar, which can be seen in the lower graph of Figure \ref{fig:exe}. This relative load can be read as the number of rotations needed to completely update $\L$. In this case it is about $2.3$ rotations since the worst case sector is decisive. If the rotation rate is adjustable it can be decreased to match a lower worst case value or increased to match a higher worst case value. 
\par 

\begin{figure}[t]
	\includegraphics[width=0.475\textwidth]{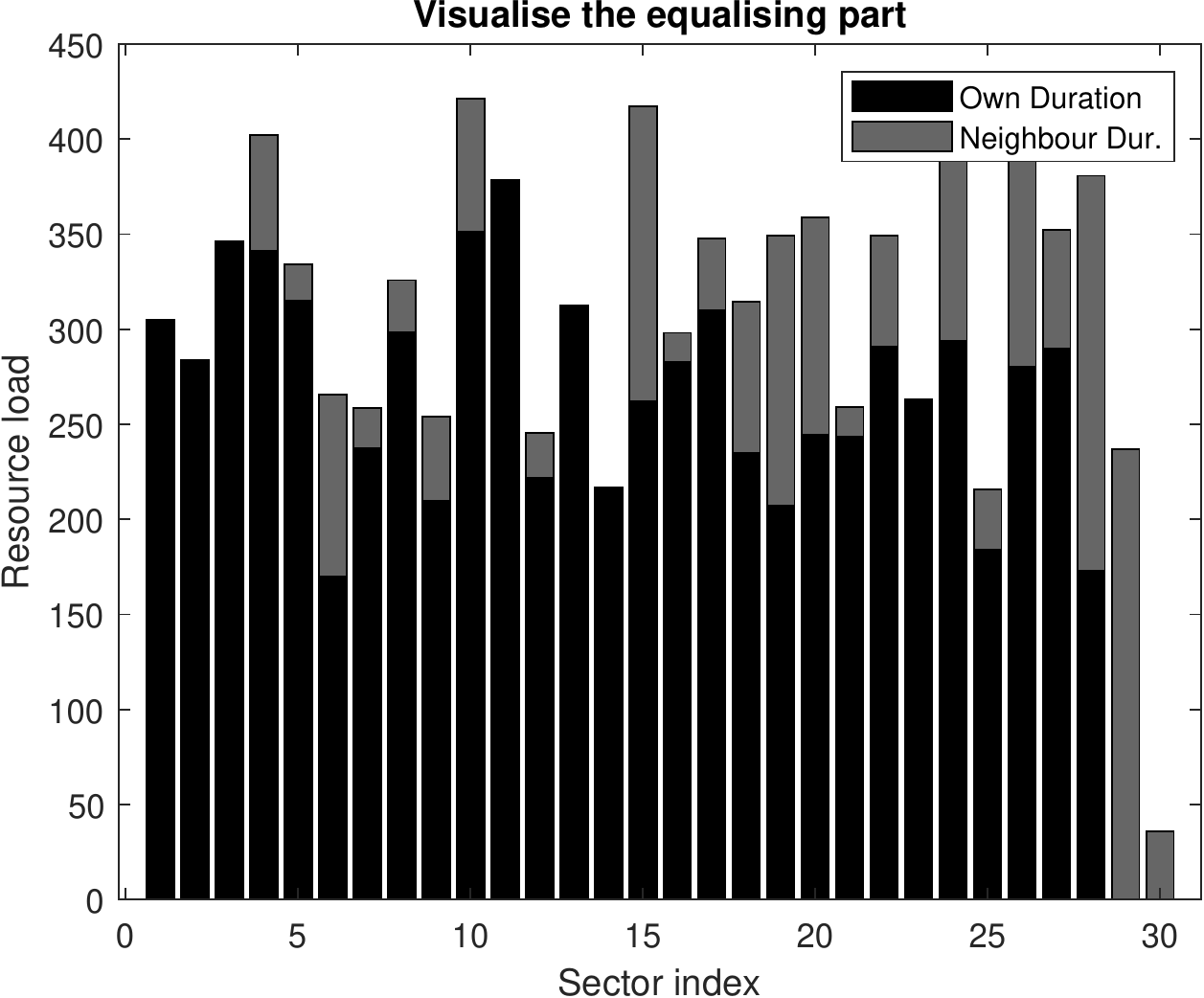}\hfill
	\includegraphics[width=0.475\textwidth]{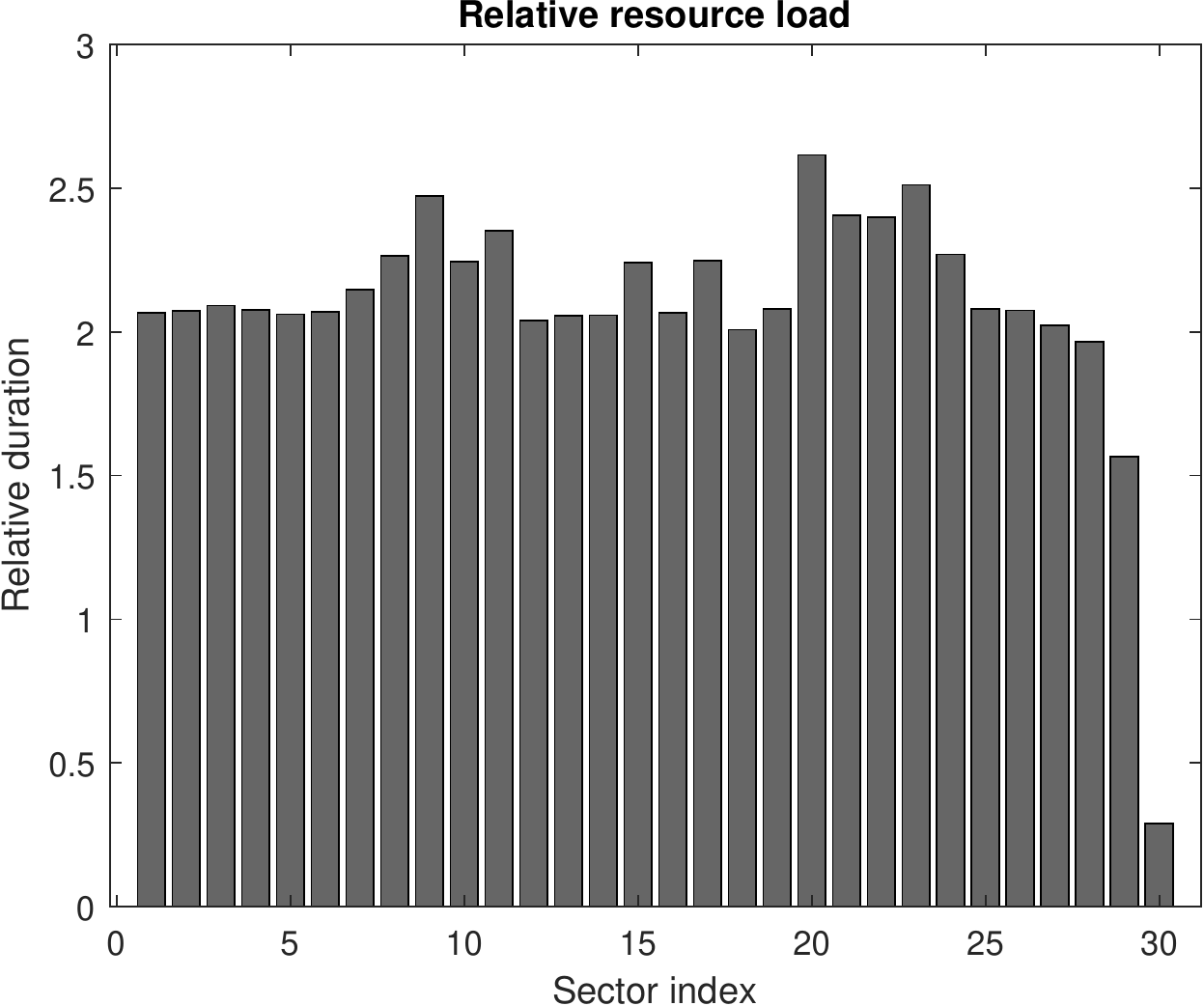}
	\caption{These graphics show the result for the same experiment as in Figure \ref{fig:ex}, but with a field of view defined by $n=1$.\label{fig:exz}}
\end{figure}

If this is not possible the SNR guarantee in every update step can be increased by decreasing the field of view. Figure \ref{fig:exz} shows the same experiment as before but with a field of view of $3$ sectors ($n=1$) which is about $36^\circ$. As expected the resulting performance decreases a bit. The advantage is, that the steering losses decrease as well. This leads to a simple method in either adapting the rotation rate to the available performances or adapting the usage of the resources to the rotation rate.

\section{Algorithm with other scheduling requirements} \label{sec:othercases}
In many cases the radar resource management does the resource allocation for the surveillance tasks dynamically. This contradicts to the assumption that this is an input parameter for the algorithm. For dynamic allocation, as it is done for instance by a priority-based scheduler \cite{butlerP}, the allocation is measured over time. Under the assumption that the allocation does not change drastically the result of the proposed algorithm still can give a usable scheme. The disadvantage is that it depends on the worst guess. If for a sector the available resource was estimated two times the actual available resource the revisit time will be approximately two times worse than estimated. An additional online algorithm for fast adaptation will be presented in the future. 
\par

\section{Conclusion and further work} \label{sec:conclusion}
In this paper a simple surveillance algorithm for multifunctional radar systems that equalises the revisit time of surveillance tasks for all beam pointing directions is presented. The goal was to decouple resources and their corresponding revisit times per sector. To achieve this, a sector which has resources above the average, can support other sectors that are occupied with other tasks such as track updates or classification measurements. An additional advantage is the fast computability due to greedy approaches. An easy adaptation to existing scheduling schemes is possible such that even existing radar systems could be upgraded.
\par
In the future the adaptation due to sudden and drastic changes in the environment will be improved. This may be done by intelligent sorting of the update tasks within each sector. It was mentioned that the rotation rate can be adapted to the dynamic surveillance scheme. In a next step a realistic optimising condition for this will be investigated as well. Additionally a comparison to earliest deadline first will be conducted.

\printbibliography

\end{document}